\documentclass[nofootinbib,floatfix,superscriptaddress,a4paper,
twocolumn,PRD]{revtex4-1}
\usepackage{amsmath}
\usepackage{amssymb}
\usepackage{graphicx}
 \usepackage{color}
\usepackage[T1]{fontenc} 
\usepackage{slashed}
\usepackage{ulem}
\def\bpm{\begin{pmatrix}} 
\def\epm{\end{pmatrix}} 
\def\bea{\begin{eqnarray}}
\def\eea{\end{eqnarray}}

\definecolor{mkgreen}{rgb}{0.2,.70,.3}

\newcommand{\fig}[1]{Fig.~\ref{#1}}

\newcommand{\lamR}{\lambda_R}
\newcommand{\lamS}{\lambda_S}

\newcommand{\vs}{v_S}

\def\lsim{\raise0.3ex\hbox{$\;<$\kern-0.75em\raise-1.1ex\hbox{$\sim\;$}}}
    
\usepackage[pdftex,colorlinks,linkcolor=blue,citecolor=red,urlcolor=blue]{hyperref} 
\hypersetup{linktocpage}

\newcommand{\AddrWue}{%
Institut f\"ur Theoretische Physik und Astronomie, 
Universit\"at W\"urzburg\\
97074 W\"urzburg, Germany}

\allowdisplaybreaks

\begin{document}

\title{Is the CMS $eejj$ excess a hint for light supersymmetry?}

\author{Manuel E. Krauss} \email{manuel.krauss@physik.uni-wuerzburg.de}
\author{Werner Porod} \email{porod@physik.uni-wuerzburg.de}
\affiliation{\AddrWue}

\begin{abstract}
We discuss the impact of additional
 two-body decays of the right-handed neutrino into a light charged Higgs 
state on the dilepton plus dijet cross sections from resonant $W'$ production. 
We consider in particular a supersymmetric left-right symmetric model which predicts
such a light charged Higgs boson. We demonstrate that the $eejj$ excess as measured by CMS can be explained
best if the $W'$ also has decay modes into Higgsino-like charginos and neutralinos
with masses of a few hundred GeV. Provided that this excess is confirmed, the model
predicts also one right-handed neutrino with a mass below 200 GeV as well as a doubly charged
Higgs boson which should be discovered at the LHC in the near future.
\end{abstract}

\maketitle

 \section{Introduction}

The Standard Model of particle physics (SM) has so far persisted every test at current colliders, and 
after the discovery of a particle consistent with the SM Higgs boson \cite{ATLAS:2012ae,Chatrchyan:2012tx}, 
for the first time 
a complete theory of particle physics exists.
Nevertheless, the SM lacks explanations for several phenomena such as neutrino masses or dark matter 
and in addition inherits a big naturalness problem so that modifications are
called for. Supersymmetry (SUSY) is among the most promising candidates for physics beyond the SM as it
relaxes the hierarchy problem and provides a candidate for dark matter. The minimal
supersymmetric standard model  is currently 
under pressure
as it prefers 
smaller Higgs masses than the measured 125~GeV and moreover also provides no mechanism for neutrino mass generation.
The constrained minimal
supersymmetric standard model can already be excluded at the 90\% confidence level \cite{Bechtle:2015nua}.

In the presence of extended gauge groups, as is the case in left-right (LR) symmetric theories, 
those issues are naturally resolved. 
In this class of models, the tree-level Higgs boson mass gets enhanced
if the larger gauge group is broken to the SM group near the TeV scale, see, e.g., 
 Refs.~\cite{Babu:1987kp,Zhang:2008jm,Hirsch:2011hg,Krauss:2013jva} and references therein, and right-handed neutrinos ($\nu_R$)
provide the basis for the generation of light neutrino masses by a seesaw mechanism.
LR symmetry relies on the gauge group $SU(3)_c \times SU(2)_L\times SU(2)_R \times U(1)_{B-L}$
 and can be thought of as the remnant of a broken $SO(10)$ or Pati Salam gauge group.
Those models predict, besides the presence of three $\nu_R$ states, a plethora of new particles, among them, several neutral and charged Higgs fields in association with
the different gauge symmetry breaking steps as well as extra gauge bosons.

Recently, in a search for an extra charged gauge boson ($W'$) and right-handed neutrinos ($\nu_R$), a 
$2.8~\sigma$ excess was observed  at the CMS experiment \cite{Khachatryan:2014dka}, whereas
the cross section is too low when compared to the prediction of simplified LR models. 
It has, however, been shown that it can
 be explained in LR symmetric frameworks when departing from the assumption that 
 the $SU(2)_{L,R}$ gauge couplings are of the same strength \cite{Deppisch:2014qpa,Heikinheimo:2014tba},
or when including neutrino mixing and $CP$ phases \cite{Aguilar-Saavedra:2014ola,Gluza:2015goa}.
In the context of other models, it has been shown that resonant  production of sleptons in $R$-parity violating SUSY \cite{Allanach:2014nna,Allanach:2014lca} and 
of vectorlike leptons \cite{Dobrescu:2014esa} can fit the excess.
Implications for leptogenesis have been explored in Refs.~\cite{Dhuria:2015wwa, Dhuria:2015hta}.

In this paper, we show how the necessary cross section is obtained in left-right supersymmetric models
featuring in the Higgs sector two $SU(2)$ bidoublets and two $SU(2)_R$ triplets. 
In this model a light charged Higgs is predicted into which the 
 right-handed neutrino can decay \cite{Brooijmans:2014eja,Basso:2015pka}. 
 By this effect, the cross section of the sought-after final state is reduced by the 
right amount. \footnote{This possibility was already qualitatively noted in Ref.~\cite{Basso:2015pka} and shall 
be quantitatively proven here.}

\section{Model}

We consider a minimal realization of left-right supersymmetry where the breaking of 
left-right symmetry works via Higgs fields in the adjoint representation of $SU(2)_R$. These are 
$\Delta_{1R},~\Delta_{2R}$, transforming as (\textbf{1},\textbf{1},\textbf{3},$\mp$2) under 
$SU(3)_c \times SU(2)_L \times SU(2)_R \times U(1)_{B-L}$.
For a nontrivial Cabibbo-Kobayashi-Maskawa matrix, two Higgs bidoublets 
$\Phi_1$, $\Phi_2$, which transform as (\textbf{1},\textbf{2},\textbf{2}$^*$,0), are required. 
In addition, a gauge singlet superfield $S$, which guarantees the breaking of 
left-right symmetry in the supersymmetric limit, is introduced.
The superpotential reads in its most general form
\begin{align}
   &W =\nonumber
   y^Q_1\, \hat Q_L \cdot \hat \Phi_1 \cdot \hat Q_R +
   y^Q_2\, \hat Q_L \cdot \hat \Phi_2 \cdot \hat Q_R +
   y^L_1\,  \hat L_L \cdot \hat \Phi_1 \cdot \hat L_R \\ \nonumber   &~+
    y^L_2\, \hat L_L \cdot \hat \Phi_2 \cdot \hat L_R + 
   y^L_3\,  \hat L_L \cdot \hat \Delta_{2L} \cdot \hat L_L + 
   y^L_4\, \hat L_R \cdot \hat \Delta_{1R} \cdot \hat L_R \\ \nonumber 
   &~+ \Big( \mu_L + \lambda_L \hat S \Big)\hat \Delta_{1L} \cdot \hat \Delta_{2L} 
   + \Big( \mu_R + \lamR \hat S \Big)\hat \Delta_{1R} \cdot \hat \Delta_{2R} \nonumber \\ 
   &~+ \Big( \mu_1 + \lambda_1 \hat S \Big)\hat \Phi_1\cdot \hat \Phi_1
   + \Big( \mu_2 + \lambda_2 \hat S \Big)\hat \Phi_2\cdot \hat \Phi_2 \nonumber \\ &~+ \Big( \mu_{12} + \lambda_{12} \hat S \Big)\hat \Phi_1 \cdot \hat \Phi_2
   + \frac13 \lamS \hat S^3 + \mu_S \hat S^2 + \xi_S \hat S\,,
\label{eq:Wtrip}
\end{align}
and we refer to Ref.~\cite{Alloul:2013fra} for the detailed index structure.
All superfields with the respective quantum numbers are defined in Table~\ref{tab:matter_content}.

\begin{table}[htbp]
\centering
\begin{tabular}{|c|c|c|c|} 
\hline \hline 
{\small{Superfield}} & {\small{Spin 0}} & {\small{Spin \(\frac{1}{2}\)}}&
{\small{quantum numbers}}\\ 
\hline \hline
\(\hat{Q}_L = (\hat u_L,\hat d_L)\) & \(\tilde{Q}_L\) & \(Q_L\) & \(({\bf 3},{\bf 2},{\bf 1}, \frac{1}{3}) \)\\ 
\({\hat{Q}_R} = (\hat u_R^c,\hat d_R^c)\) & \(\tilde{Q}_R\) & \(Q_R\) &\(({\bf \overline{3}},{\bf 1},{\bf 2^*},-\frac{1}{3}) \)\\ 
\(\hat{L}_L= (\hat \nu_L,\hat \ell_L)\) & \(\tilde{L}_L\) & \(L_L\) &  \(({\bf 1},{\bf 2},{\bf 1},-1) \) \\ 
\({\hat{L}_R}= (\hat \nu_R^c,\hat \ell_R^c)\) & \(\tilde{L}_R\) & \(L_R\) & \(({\bf 1},{\bf 1},{\bf 2^*},1) \)\\ 
\(\hat{\Phi}_i\) & \(\Phi_i\) & \(\tilde{\Phi}_i\) & \(({\bf 1},{\bf 2},{\bf 2^*},0)\) \\ 
\(\hat{\Delta}_{1L}\) & \(\Delta_{1L}\) & \(\tilde{\Delta}_{1L}\) &  \(({\bf 1},{\bf 3},{\bf 1},-2)\) \\ 
\(\hat{\Delta}_{2L}\) & \(\Delta_{2L}\) & \(\tilde{\Delta}_{2L}\) &  \(({\bf 1},{\bf 3},{\bf 1},2)\) \\ 
\(\hat{\Delta}_{1R}\) & \(\Delta_{1R}\) & \(\tilde{\Delta}_{1R}\) &  \(({\bf 1},{\bf 1},{\bf 3},-2)\) \\ 
\(\hat{\Delta}_{2R}\) & \(\Delta_{2R}\) & \(\tilde{\Delta}_{2R}\) &  \(({\bf 1},{\bf 1},{\bf 3},2)\) \\ 
 \({\hat{S}}\) & \(S\) & \( \tilde S\) &  \(({\bf 1},{\bf 1},{\bf 1},0) \)\\ 
\hline \hline
\end{tabular} 
\caption{Chiral superfields and their quantum numbers with
respect to \( SU(3)_c\times\, SU(2)_L\times\,  SU(2)_R\times\,
U(1)_{B-L}\). }
\label{tab:matter_content}
\end{table}

The gauge group $SU(2)_R \times U(1)_{B-L}$ gets broken down to the hypercharge group $U(1)_Y$ 
once the neutral components of the
$SU(2)_R$ triplets 
develop vacuum expectation values (vevs) according to 
\begin{align}
\langle \Delta_{1R} \rangle &= 
\Big\langle \begin{pmatrix}
\frac{\Delta_{1R}^-}{\sqrt{2}} & \Delta_{1R}^0 \\
\Delta_{1R}^{--} & -\frac{\Delta_{1R}^-}{\sqrt{2}}
\end{pmatrix}\Big\rangle = 
\begin{pmatrix}
0 & \frac{v_{1R}}{\sqrt{2}} \\
0 & 0
\end{pmatrix}\,, \nonumber \\
\langle \Delta_{2R} \rangle &=
\Big\langle \begin{pmatrix}
\frac{\Delta_{2R}^+}{\sqrt{2}} & \Delta_{2R}^{++} \\
\Delta_{2R}^{0} & -\frac{\Delta_{2R}^+}{\sqrt{2}}
\end{pmatrix}\Big\rangle = 
\begin{pmatrix}
0 & 0\\
\frac{v_{2R}}{\sqrt{2}}  & 0
\end{pmatrix}\,.
\label{eq:desired_vacuum}
\end{align}
For further reference we define $v_{1R}^2 + v_{2R}^2 = v_R^2,~\tan \beta_R = v_{2R}/v_{1R}$.
Note that at tree-level
this does not correspond to the true vacuum \cite{Kuchimanchi:1993jg,Huitu:1994zm}. Instead, a configuration 
where the vevs are aligned along the $\sigma_1$ direction ($\sigma_1$ being the first Pauli matrix) corresponds to the 
global minimum, hence bearing vevs for the doubly charged fields $\Delta_{1R}^{--}$ as well as 
 $\Delta_{2R}^{++}$ which breaks 
the electromagnetic $U(1)$.
The desired vacuum structure of Eq.~(\ref{eq:desired_vacuum}), in turn, 
only features a saddle point at tree level which translates to a tachyonic lightest
doubly charged Higgs field $H^{\pm \pm}_1$. 
At the one-loop level, however, this situation 
changes \cite{Babu:2008ep,Babu:2014vba,Basso:2015pka}. 
In Ref.~\cite{Basso:2015pka} it was shown that $H^{\pm \pm}_1$ receives masses at the order 
of a few hundred GeV while the desired, charge-conserving vacuum can indeed be the
global minimum at one loop.

Neutrinos get their masses from different sources: the $SU(2)_R \times U(1)_{B-L}$ breaking induces
 Majorana masses for the right-handed neutrinos. The final $SU(2)_L \times U(1)_{Y}$ breaking provides
two sources for the masses of left-handed neutrinos: 
(i) seesaw type II due to the $SU(2)_L$ triplets which have the corresponding vev structure as the $SU(2)_R$ counterparts, and
(ii) seesaw type I contributions due to the bidoublets which receive vevs according to 
\begin{align}
\langle \Phi_1 \rangle &=\Big\langle 
\begin{pmatrix}
\Phi_1^0 & \Phi_1^+ \\
\Phi_1^- & \Phi'{}_1^{0}
\end{pmatrix} \Big\rangle = 
\begin{pmatrix}
\frac{v_d}{\sqrt{2}} & 0 \\
0 & \frac{v_1'}{\sqrt{2}}
\end{pmatrix} \,, \nonumber \\ 
\langle \Phi_2 \rangle &=\Big\langle
\begin{pmatrix}
\Phi'{}_2^0 & \Phi_2^+ \\
\Phi_2^- & \Phi_2^{0}
\end{pmatrix} \Big\rangle = 
\begin{pmatrix}
\frac{v_2'}{\sqrt{2}} & 0 \\
0 & \frac{v_u}{\sqrt{2}}
\end{pmatrix}\,.
\label{eq:EW_vevs}
\end{align}
The vevs $v_i'$ give rise to $W-W'$ mixing and are constrained by kaon data so that
we will neglect them in what follows.
The $W-W'$ mixing induced by these vevs has been exploited by Refs.~\cite{Brehmer:2015cia,Dobrescu:2015yba} to explain
the recent excess in diboson events seen by ATLAS \cite{Aad:2015owa}.
For further reference we define 
$v_d^2 + v_u^2 = v^2,~\tan \beta = v_u/v_d$.

The singlet
 dynamically generates effective $\mu$ terms by acquiring a vev,
\begin{align}
\langle S \rangle = \frac{\vs}{\sqrt{2}} \,.
\label{eq:singlet_vev}
\end{align}

In the following, we assume a discrete $\mathbb{Z}_3$ symmetry under which scalar fields transform as 
$\phi \to e^{\frac{2 \pi i}{3}} \phi$ and which consequently forbids all bilinear and linear terms in the superpotential,
 $\mu_i  = \xi_S =  0$, and in addition we assume $\lambda_{1} = \lambda_2 = 0$, analogously to
Refs.~\cite{Basso:2015pka,Alloul:2013fra, Brooijmans:2014eja}. 
The effective $\mu$ terms are then given by
\begin{align}
\mu_{\rm eff} = \frac{\lambda_{12} v_S}{\sqrt{2}}\,,~\mu_{R,{\rm eff}} = \frac{\lambda_R v_S}{\sqrt{2}}\,.
\end{align}

For simplicity we assume all vevs and couplings apart from the
quark Yukawa couplings to be real and for phenomenological reasons 
 the hierarchy 
$v_S, v_{1R,2R} \gg v_{u,d} \gg v_{1,2}',v_{1L,2L} \approx 0 $.

We remark that in this class of models the neutrinos are Majorana particles. This can in general not 
explain the reported discrepancy between dilepton final states of the same sign and of different signs 
in the $\ell \ell j j$ searches in Ref.~\cite{Khachatryan:2014dka} 
\footnote{This mismatch is fortified by the fact that an ATLAS analysis searching for same-sign dileptons and two jets sees no excess over the background \cite{Aad:2015xaa}.}
 as a Majorana $\nu_R$ state decays
equally into $\ell^+ +X^-$ and $\ell^- +X^+$ if $CP$ is  conserved in the lepton sector. 
For the explanation of the deficit
of the same-sign dilepton final state one could use the inverse seesaw mechanism to obtain quasi-Dirac
neutrinos without changing the qualitative results for the signal.
An alternative explanation has been provided in Ref.~\cite{Gluza:2015goa} where the interference of
two right-handed Majorana neutrinos with mixed flavor content and opposite $CP$ parities can partially 
suppress same-sign lepton pairs in the considered process.

\section{Mass spectrum}
Here, we briefly review the relevant particles and masses for the subsequent discussion:
\begin{itemize}
\item[(i)] The most obvious consequence of models with an extended gauge sector is the presence of 
extra massive gauge bosons. Left-right symmetric models
feature heavy $W'^\pm$ bosons as well as a neutral $Z'$. Because of the breaking via Higgs triplets,
 the $W'$ turns out to be lighter than the $Z'$,
\begin{align}
M_{W'} \simeq \frac{g_R}{\sqrt{2}} \,v_R\,\text{ and }
M_{Z'} \simeq \sqrt{g_{R}^2 + g_{BL}^2} \, v_R\,,
\end{align} 
with $g_R$ and $g_{BL}$ being the gauge couplings associated with the $SU(2)_R$ and $U(1)_{B-L}$ gauge groups.
Thus, the searches for a $W'$ give the stronger constraints in this class of models.
The strongest bounds stem from  searches for hadronically decaying resonances. 
The searches in the $tb$ final state exclude resonances below roughly 2~TeV \cite{Chatrchyan:2014koa}. The exact bound depends on
the parameters and the width of the $W'$ and reads $M_{W'} \leq 1.95~$TeV in case of a large width due to
many $W'$ decay modes \cite{Basso:2015pka}. The dijet bounds \cite{Khachatryan:2015sja} depend even 
more strongly on the decay width. The reason is that a small excess of events at an invariant mass of 
1.8~TeV has been observed so that the cross section limits in its vicinity are weaker. In 
\fig{fig:qqbounds}, we display the experimental
result (red line) and the predicted signal cross section (gray band). The width of the gray band
is given by the fact that, depending on the model, additional $W'$ decay channels open. The upper edge
corresponds to the case that only decays into SM particles and $\nu_R$ are open, whereas the lower edge
corresponds to the case that also light supersymmetric particles are present in the $W'$ final states.
This search hence only excludes masses below
$\sim$ 1.75~TeV for a light supersymmetric mass spectrum and for a heavy one masses below $\sim$ 2.25~TeV. 
\begin{figure}
\includegraphics[width=.9\linewidth]{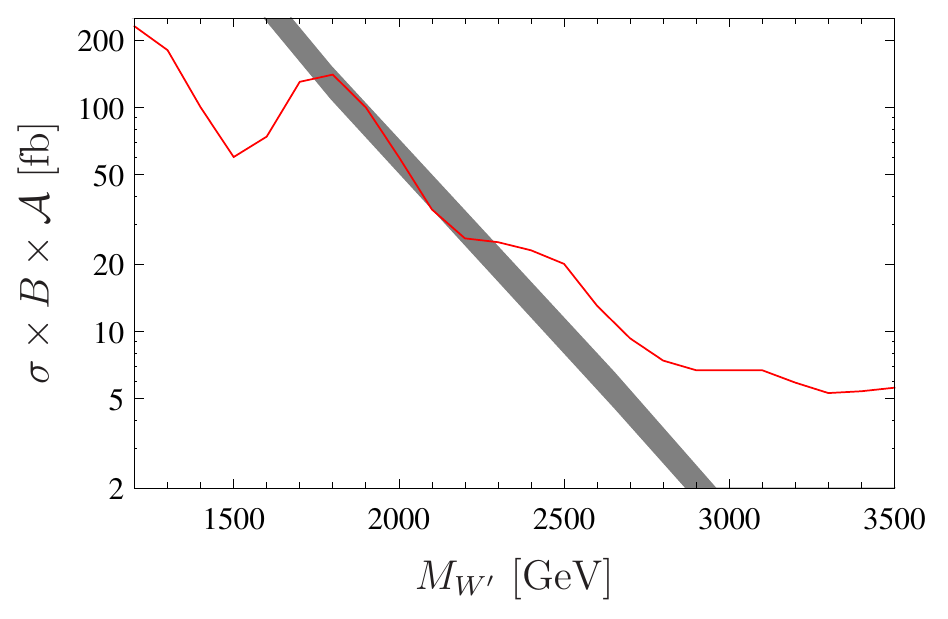}
\caption{Comparison of the cross section of a resonantly produced $W'$ decaying into a dijet final state,
 $\sigma(W') \times {\rm BR}(W' \to j j) \times \mathcal A$ (gray band), with the observed limits from Ref.~\cite{Khachatryan:2015sja} (red line) at $\sqrt{s}=8~$TeV, using $g_R=g_L$. $\mathcal A\approx 0.53 $ is the acceptance
corresponding to   the kinematic cuts used in the analysis. The leading-order results as obtained from {\tt MadGraph} 
are normalized to the next-to-leading order by applying a
 $K$ factor  1.23 as in Ref.~\cite{Khachatryan:2015sja}. The gray band gives
the allowed range for this cross section varying over the model parameters, in particular $ 150~{\rm GeV}< \mu_{\rm eff} < 1.8~{\rm TeV}$.
 }
\label{fig:qqbounds}
\end{figure}
Assuming a discrete charge-conjugation or parity invariance of the Lagrangian, 
the constraints from flavor-changing neutral currents in meson systems can give much tighter lower bounds  on minimal nonsupersymmetric 
 left-right models of roughly $M_{W'} \gtrsim 3$~TeV while requiring the additional Higgs states to be in the multi-TeV region \cite{Bertolini:2014sua,Maiezza:2010ic}. 
 However, those constraints can already be evaded in nonsupersymmetric left-right scenarios 
 \cite{Guadagnoli:2010sd} and are even more relaxed in supersymmetric models due to the competing SUSY 
 box diagrams, so in the model under consideration, the constraints are lowered to
 $M_{W'} \gtrsim 2~$TeV  even if parity symmetry is assumed \cite{Zhang:2007qma} while also the lighter 
 Higgs states survive the constraints \cite{Babu:2008ep}.
\item[(ii)] As discussed before, neutrino masses are generated by the Majorana neutrino Yukawa coupling $y^L_4$.
Using $y_2^L \, v_u \ll y^L_4 \, v_{1R}$, the masses of the right-handed neutrinos are given by
\begin{align}
 m_{\nu_R} \simeq \sqrt{2} \, y^L_4 \, v_{1R}\,.
\end{align}
For $v_{1R}$ at the TeV scale and $y_L^4$ of $\mathcal O(1)$, one needs $|y_2^L|_{ij} \lsim \mathcal O(10^{-5})$ to explain the light neutrino masses.
\item[(iii)] Their superpartners, the right sneutrinos,  receive contributions to their mass matrix by the 
soft SUSY breaking masses $m_{L_R}^2$ as well as $D$ and $F$ terms. The $F_{\Delta_{1R}}$ term 
 splits the right sneutrinos into their scalar (S) and pseudoscalar (P) parts. Their
masses read, in the limit of vanishing $A_{y^L_4}$ terms and vanishing left-right mixing in the sneutrino sector, as
\begin{align}
(m_{\tilde \nu_R}^{S/P})^2 = &m_{L_R}^2 + \frac{1}{4} (g_R^2+g_{BL}^2) (v_{2R}^2-v_{1R}^2)  \notag \\
&+ y_4^L (2\, y_4^L\, v_{1R}^2 \pm \lambda_R\, v_S\, v_{2R}) \,. \label{eq:sneutrino_mass}
\end{align}
Large values for $\mu_{R,{\rm eff}}$ lead to a considerable mass splitting between the $CP$ eigenstates
and are hence constrained by the requirement that $\tilde \nu_R^{S/P}$ does not get 
tachyonic. 
Moreover, this splitting can lead to the situation in which some sneutrino eigenstates are light, whereas all charged 
sleptons are much heavier.
\item[(iv)] As mentioned above, the lightest doubly charged Higgs in the $SU(2)_R$ sector gets its
mass radiatively.
Necessary for that to happen is small $\tan \beta_R$ in the range   $1.01 - 1.05$.
Large Majorana neutrino couplings $y^L_4$ have the effect of decreasing $m_{H^{\pm\pm}_1}$, but this 
can be compensated by $\mu_{R,{\rm eff}}$ \cite{Basso:2015pka}. Depending on whether it
decays into $e e$, $\mu \mu$, and $\tau \tau$, the bounds on its mass are 
444, 459, and 204~GeV, respectively \cite{Chatrchyan:2012ya}.

\item[(v)] As apparent from the Higgs representations , there are 
six physical singly charged Higgs bosons.
The mostly $\Phi_2^-$-like state is usually the lightest out of the six states. Its mass can be 
approximated by
\begin{align} 
m_{H^\pm_1}^2 \simeq 
g_R^2 \, v_R^2\, \frac{\tan^2 \beta_R -1}{2(1+\tan^2 \beta_R)}\,.
\label{eq:Hpm_mass}
\end{align}
Moreover, a small mixing with the $\Delta_{1R}^-,\Delta_{2R}^+$ state exists due to the 
$D$-term contributions. The $\Delta_{1R}^-$ 
component within $H^\pm_1$ is roughly
\begin{align}
\mathcal R_{H^-_1,\Delta_{1R}^-} \simeq \frac{v}{2\, v_R}\,.
\end{align}
This admixture is responsible for the $\nu_R$-$H^\pm_1$-$\ell^\mp$ coupling   which
is $\propto y^L_4 \cdot \mathcal R_{H^-_1,\Delta_{1R}^-}$. 
We have checked that the approximate formulas   
agree with the full numerical results within  5~\%.
\end{itemize}

\section{Explaining the $eejj$ excess}

We now turn to the interpretation of the analysis of $\ell \ell j j $ final states from
Ref.~\cite{Khachatryan:2014dka}.
It has been noted that in the $W'$ mass window $1.8 \lsim M_{W'}/{\rm TeV} \lsim 2.4$ a 2.8~$\sigma$ excess in
the $e e j j$ final state has been observed whereas nothing has been seen in the muonic channel.
It has further been noted that the signal does not appear to be consistent with the 
assumption of a $W'$ from a left-right-symmetric model which decays into $\ell \nu_R$. The main reason is that 
the expected production cross section of said final state while assuming left-right symmetry is too large
by a factor of roughly 3--4.

\subsection{Ways to reduce the signal cross section}

Let us recall the assumptions on which the corresponding bounds
on $W'$ and $\nu_R$ masses are based. First, it is assumed that the gauge coupling strengths
$g_L$ and $g_R$ are of equal size. We shall also stick to that since left-right symmetry and also $D$ parity
are broken near the TeV scale. An analysis where $D$ parity is broken at the high scale with the 
resulting $g_R < g_L$ at $M_{W'}$ has been performed in Refs.~\cite{Deppisch:2014qpa,Heikinheimo:2014tba}. 
As a consequence of taking $g_L = g_R$, we also have to include the $W'$ bounds from $t b$ searches 
so that the region of interest shrinks to $1.95 \lsim M_{W'}/{\rm TeV} \lsim 2.4$.
Furthermore, the experimental analysis considers no off-diagonal couplings of right-handed neutrinos which would eventually
lead to $\ell_i \ell_k j j$, $i \neq k$. The effect of that has, e.g., been explored in Ref.~\cite{Aguilar-Saavedra:2014ola}, and we shall keep diagonal $y^L_4$ \footnote{Note that there is still enough
freedom in the other neutrino Yukawa couplings $y^L_i$, in particular when allowing non-zero
$v_k'$ and $v_{kL}$, to explain light neutrino data.}. 

Moreover, Ref.~\cite{Khachatryan:2014dka} uses a simplified model where only a $W'$ as well as 
up to three generations of right-handed neutrinos are added with respect to the Standard Model field
content. As a consequence, the $W'$ only decays into $q \bar q$ and $\ell \nu_R$ final states whereas
the dominating possibility for $\nu_R$ to decay further is a three-body decay via a virtual $W'^*$:
\begin{align}
p p \to W' \to \ell \,\nu_R  \to \ell\, W'^* \ell \to \ell\, \ell\, j\, j\,.
\end{align}
One has to remain aware of this when interpreting the respective bounds, particularly since 
the Higgs sector needs to be enlarged with respect to the Standard Model 
when constructing a left-right symmetric model. 
In our  setup, we find that, in particular, the assumption that 
only three-body decay channels are open is in general not correct; the light 
$H^\pm_1$ state with non-negligible couplings to $\nu_R \, \ell$ turns out to 
be a viable final state, resulting in
\begin{align}
\nu_R \to  H^\pm \, \ell^\mp
\end{align}
in most of the parameter space.\footnote{The effect of a light charged Higgs on such observables
has already been noted in Ref.~\cite{BarShalom:2008gt}.} As long as it is kinematically available, $H^\pm_1$ will further
decay into $t b$, and hence not or only marginally contribute to the searched-for final state. 
The relative importance of the two-body decay vs the three-body decay modes depends mainly
on two features: (i) the size of the small admixture of $\Delta_{1R}^-$ within $H_1^-$ and (ii)
the ratio $m_{\nu_R}/M_{W'}$ as this gives the propagator suppression of the three-body decays.
However, a significant reduction of the $\ell \ell j j $ final state by 
several tens of percent can in general be the case.

Furthermore, in general, the $W'$ decays into $q\bar q / \ell\nu_R$ do not exhaust all possibilities:
decays into SUSY particles, e.g., into charginos $\tilde \chi^\pm$ and  neutralinos $\tilde \chi^0$
or into sleptons/squarks or into vector and Higgs bosons, e.g., $W$ and $H^0$ or $H^\pm$ and $H^{\mp\mp}$,
are as well possible. 
This leads of course to a broadening of the $W'$ width, thereby reducing the branching fraction into
the above considered final states. More importantly, if the sum of the masses of an additional $W'$ final
 state does not exceed $m_{\nu_R}$, this will also be a viable final state of the three-body decay:
 \begin{align}
 \nu_R \to \ell \, W'^* \to \ell \, X\,Y\,,~~X,Y\neq jj\,.
 \end{align}
 
\begin{figure}
\includegraphics[width=.9\linewidth]{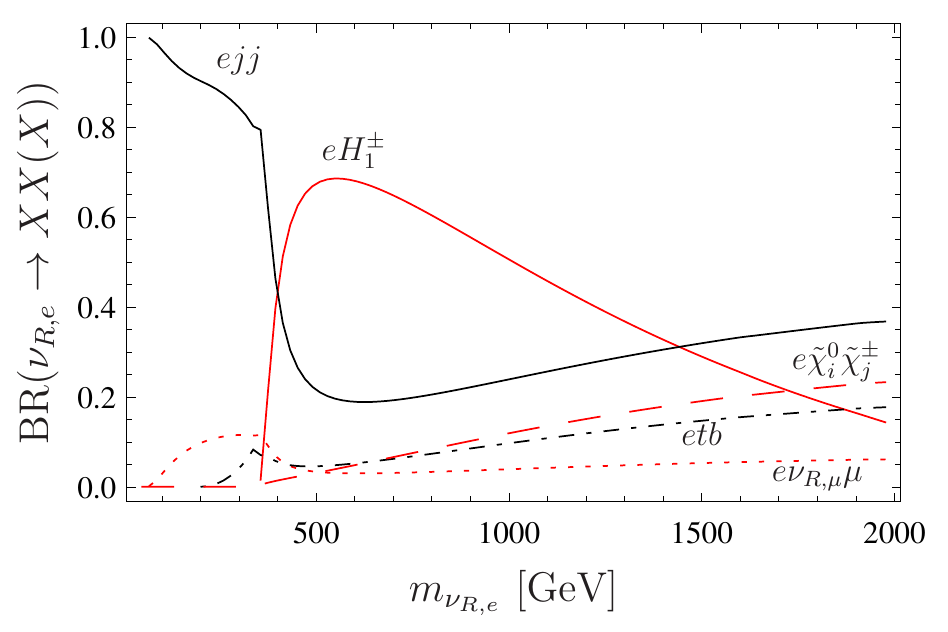}
\caption{Branching ratios of the $\nu_{R,e}$ as a function of its mass assuming $v_R=5~$TeV,
$\tan \beta_R=1.02$ and $\mu_{\rm eff}=150~$GeV as well as $m_{\nu_R,\mu}=50~$GeV and $m_{\mu_{R,\tau}}=2.5~$TeV. The decay products depicted by the red solid, dashed and dotted lines are $e H^\pm_1$, $e \tilde \chi^0_i \tilde \chi^\pm_j$, 
and $e  \nu_{R,\mu} \mu$. The final states depicted by black solid and dotted-dashed lines 
are $ej j $ and $et b$ as mediated by an off-shell $W'$. }
\label{fig:nuRdecay}
\end{figure} 
 
In Fig.~\ref{fig:nuRdecay}, we exemplify this situation by showing the
branching ratios of the $\nu_{R,e}$ assuming that 
$\nu_{R,\mu}$ is lighter and is $\nu_{R,\tau}$ heavier. Because of the small $\mu_{\rm eff}$, the lightest 
chargino and neutralinos are each around 150~GeV so that they appear as decay products as well.
The sneutrinos and sleptons in this example are so heavy
that the two-body decays $\nu_R \to \tilde \nu_R\, \tilde \chi^0_i / \tilde \ell^\mp\, \tilde \chi^\pm_i $ 
are kinematically forbidden.

\subsection{Numerical results}

Let us now quantify how the above-mentioned features affect the interpretation 
of the bounds on $W'$ and $\nu_R$. For that purpose we have implemented the model into 
 {\tt SARAH} \cite{Staub:2008uz,Staub:2009bi,Staub:2010jh,Staub:2012pb,Staub:2013tta,Staub:2015kfa} and used it to create 
output for the spectrum generator {\tt SPheno} \cite{Porod:2003um,Porod:2011nf}, which allows a precise mass calculation at the one-loop order.
We have further modified the {\tt SPheno} code in order to take into account the tree-level tachyonic $H^{\pm\pm}$ and the correct calculation of its one-loop mass
as described in Ref.~\cite{Basso:2015pka}. 
Via the {\tt UFO} interface \cite{Degrande:2011ua} we have then generated model files for {\tt MadGraph} \cite{Alwall:2014hca}.
We have then calculated the relevant cross sections at the parton level and corrected them with a
suitable $K$ factor 1.3 as in Ref.~\cite{CMS:2012uwa} in order to normalize the results to the next-to-leading order.

To be consistent with the $\mu \mu j j $ search, 
the muonic $\nu_R$ must be either too heavy  or too light  in order to escape detection; 
if $m_{\nu_R}$ is almost as large as $M_{W'}$, the final state of $W' \to \nu_R \ell$ has too little phase 
space to be produced at a sizeable rate. If $\nu_R$ is very light compared to the $W'$, 
the $\nu_R$ decay products are collimated. As the reconstruction in this kind of analysis relies on high-$p_T$ objects which
are spatially well separated from each other, the signal acceptance decreases rapidly for 
$m_{\nu_R}/M_{W'} \lsim 0.1$ \cite{Khachatryan:2014dka}.

Compared with the $M_{W'} - m_{\nu_{R,\mu}}$ plane
as presented in Ref.~\cite{Basso:2015pka}, this translates to 
\begin{align}
m_{\nu_{R,\mu}}\lsim 200~{\rm GeV} ~~ {\rm or }~~  m_{\nu_{R,\mu}}\gtrsim 1.8~{\rm TeV}
\end{align}
or equivalently
\begin{align}
(y^L_4)_{22} \lsim 0.04 ~~ {\rm or} ~~(y^L_4)_{22} \gtrsim 0.36\,,
\end{align}
whereas no bounds can be set on the $\nu_R$ of tauon type.
We differentiate between three separate $\nu_R$ hierarchies: 
\begin{enumerate}
\item $m_{\nu_{R,e}} < m_{\nu_{R,\mu}} \approx  m_{\nu_{R,\tau}}$\,,
\item $m_{\nu_{R,\mu}} \ll m_{\nu_{R,e}} < m_{\nu_{R,\tau}}$\,,
\item $m_{\nu_{R,\mu}} \approx m_{\nu_{R,\tau}} \ll m_{\nu_{R,e}}$\,,
\end{enumerate}
and calculate the corresponding $ee jj $ cross sections.

The first mass hierarchy is
in fact excluded if one sets an upper bound on the the soft SUSY breaking parameter $m_{L_R}$ 
of a few TeV. It leads to a rather light $H^{++}$ state inconsistent with LHC data
\cite{Basso:2015pka}. We note for completeness that this also excludes the
hierarchy $m_{\nu_{R,\tau}} < m_{\nu_{R,e}} < m_{\nu_{R,\mu}}$. 
Hence, only the cases in which $\nu_{R,e}$ is the second-lightest or the heaviest neutrino are phenomenologically 
reasonable. 
It turns out that we can explain the observed data for both cases: a relatively light
SUSY spectrum, which is the preferred case in our model, or a rather heavy SUSY spectrum.

\subsubsection{Light supersymmetric spectrum}

\begin{figure}[htbp]
\includegraphics[width=.9\linewidth]{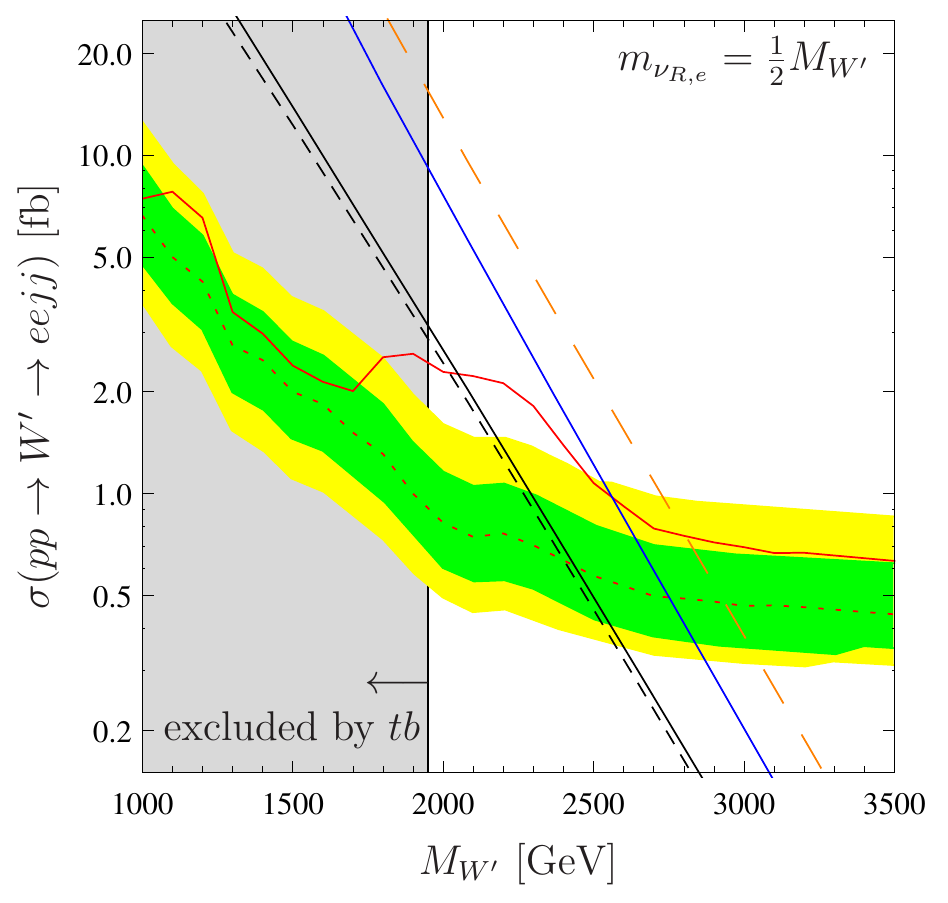}
\caption{Cross section $\sigma(p p \to W' \to e e j j)$ in fb at$\sqrt{s} = 8~$TeV for $\tan \beta_R = 1.02$ and 
$\mu_{\rm eff} = 150~{\rm GeV}$. The black solid line corresponds to the 
$\nu_R$ mass hierarchy 2 (see the text) with $\mu_{R,{\rm eff}} =4~{\rm TeV}$, whereas the black dashed line corresponds
to hierarchy 3 with $\mu_{R,{\rm eff}} = 6~{\rm TeV}$. 
The blue line corresponds to the case where 
the two-body decays $\nu_R \to \ell H^{\pm}$ have been turned off. The orange dashed curve corresponds
to a model where the SM is extended by a $W'$ and only one right-handed neutrino as used in 
Ref.~\cite{Khachatryan:2014dka}. 
The gray area is excluded from searches for resonances in $tb$ events \cite{Chatrchyan:2014koa}. 
The red solid (dotted) line corresponds to the observed (expected) exclusion limits at 95\% C.L. as given in
Ref.~\cite{Khachatryan:2014dka}, whereas the 
green (yellow) band corresponds to the expected exclusion $\pm 1\, \sigma\, (2\, \sigma)$. 
}
\label{fig:nuRbound_light}
\end{figure}

In Fig.~\ref{fig:nuRbound_light}, we show the obtained production cross section for the two phenomenologically relevant neutrino 
    mass hierarchies as black lines, using $m_{\nu_{R,e}} = 2 M_{W'}, \tan \beta_R = 1.02$ 
    and light Higgsino-like charginos and neutralinos ($\mu_{\rm eff} = 150~$GeV). 
 The gray area is excluded by $tb$ searches due to $M_{W'} < 1.95~$TeV
since the dijet bounds are weaker if light Higgsinos appear in the $W'$ final states as is the case here.
The red dotted (solid) line corresponds to the expected (observed) 95\% C.L. exclusion 
limit
as given in Ref.~\cite{Khachatryan:2014dka}. The green (yellow) band shows the expected limit
$\pm 1\,\sigma$ (2\,$\sigma$). The black solid and dashed lines 
correspond to  
the neutrino mass
hierarchies 2 and 3. 
The different values of $\mu_{R,{\rm eff}}$ for the two cases do not affect the shown cross sections but 
were adapted in order to raise $m_{H^{\pm\pm}}$ above the corresponding experimental bounds. 
The main reason for the  difference between the two $\nu_R$ mass hierarchies is the 
additional $W'$-mediated three-body decay $\nu_{R,e} \to e \, \nu_{R,\ell_i} \, \ell_i$, $i=\mu,\tau$.
We have also considered the case that the
$\nu_R$ decays only via a virtual $W'^*$ into $\ell q \bar q'$ resulting into the second-largest 
cross section shown in the blue line. Hierarchies 2 and 3 give nearly identical results
in this case, and thus only one line is shown.
For completeness, we display also the
cross section (orange line) using the same assumptions as Ref.~\cite{Khachatryan:2014dka}, namely, taking
into account only one $\nu_R$ generation within their considered simplified left-right symmetric
model. Here, we find a good agreement with their results.

From this we learn that  the 
relevant cross section  can already by reduced considerably 
if more generations of right-handed neutrinos and
 additional $W'$ decays are present. In our example, the decays of $W'$ into the light Higgsinos 
 reduces
the signal cross section by a factor $\sim 1.6$.
It is further reduced 
by a factor $\simeq 2.5$ due to the additional two- and three-body decays $\nu_{R,e} \to H^\pm_1 e^\mp$ and $\nu_{R,e} \to e^\mp \tilde \chi^0 \tilde \chi^\pm$.
The cross section
could be even further reduced if in addition light sneutrinos are present.

As apparent from Fig.~\ref{fig:nuRbound_light}, the combination of extra two-body $\nu_R$ decay channels  
and extra $W'$ channels reduces the $eejj$ cross section by the right amount so that a $W'$ 
with a mass of roughly 2~TeV could be the reason for the observed excess.

\subsubsection{Heavy supersymmetric spectrum}

We now consider the case in which all possible $W'$ decays into SUSY particles are kinematically forbidden. 
As a consequence,  the $W'$ width is smaller compared to the previous case, and the branching ratios roughly resemble 
the ones predicted for a $W'$ as considered in the simplified scenarios. As a consequence, the
dijet bounds are more important as explained beforehand in the discussion of Fig.~\ref{fig:qqbounds}, 
and thus
the $W'$ mass has to be larger than $\sim 2.25~$TeV.
This also increases the branching ratios BR($W' \to \nu_R \ell$) and BR($\nu_R \to \ell qq$)
compared to the previous case. Therefore, the $eejj$ cross section is about 1.6 times larger 
than in case of a light SUSY spectrum.

\begin{figure}[htbp]
\includegraphics[width=.9\linewidth]{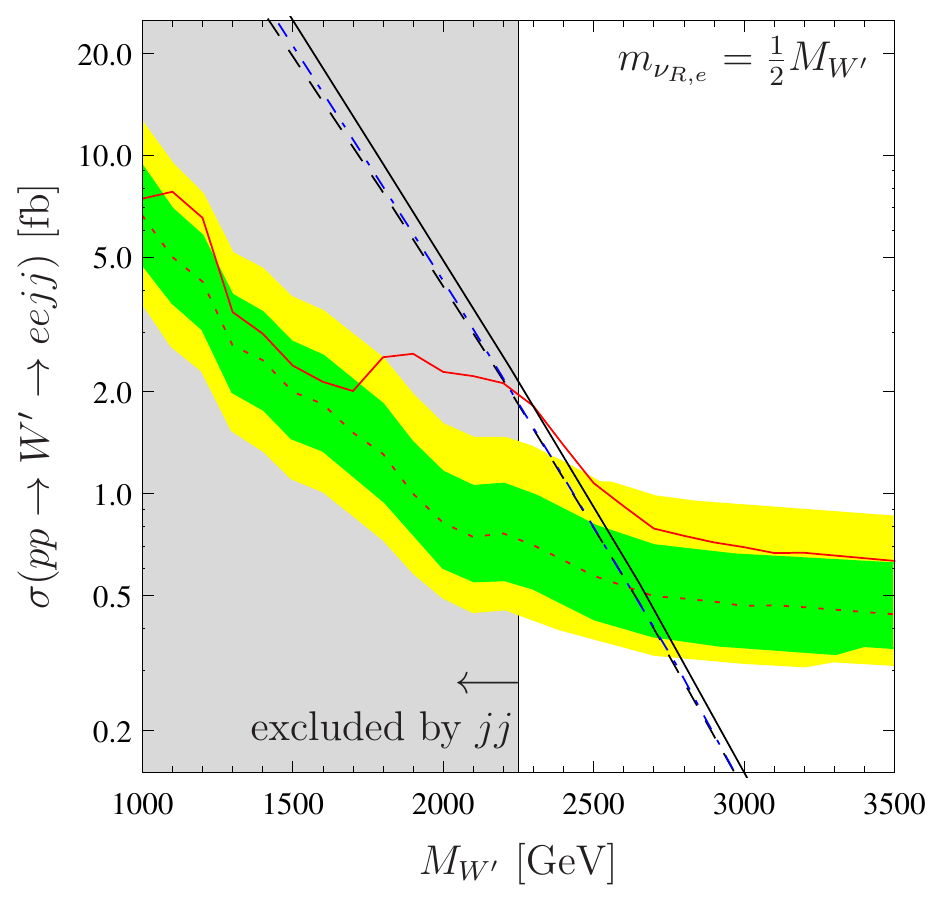}
\caption{Cross section of $p p \to W' \to e e j j $ at  $\sqrt{s} = 8~$TeV for 
$\mu_{\rm eff} = 1.5~{\rm TeV}$. The black solid (dashed) line corresponds to the 
neutrino mass hierarchy 2 with $\tan \beta_R=1.05$ (1.02) and $\mu_{R,{\rm eff}} = 6$~TeV (3.5~TeV), 
and the blue dotted-dashed line (close to the black dashed line) features
hierarchy 3 and $\tan \beta_R=1.05,~\mu_{R,{\rm eff}} = 6$~TeV. 
The gray area is excluded from searches for resonances in dijet events \cite{Khachatryan:2015sja}. 
The red solid (dotted) line corresponds to the observed (expected) exclusion limits at 95\% C.L. as given in
Ref.~\cite{Khachatryan:2014dka}, whereas the 
green (yellow) band corresponds to the expected exclusion $\pm 1\, \sigma\, (2\, \sigma)$. 
}
\label{fig:nuRbound_heavy}
\end{figure}

In Fig.~\ref{fig:nuRbound_heavy}, 
we show the resulting cross sections for mass hierarchy 2 taking  
$\tan \beta_R=1.02$ and 1.05  as well as for hierarchy 3 and $\tan \beta_R=1.05$
in the blue dotted-dashed line setting   $\mu_{\rm eff} = 1.5~$TeV.  This demonstrates that the main
effect is due to $\tan \beta_R$ as its increase leads to an increase of the the $H^\pm_1$ mass, 
see Eq.~(\ref{eq:Hpm_mass}), resulting in a decrease of BR($\nu_R \to H^+_1 \ell$).
Nevertheless, even for 
 a heavy supersymmetric spectrum, the model can explain the observed excess. 

\subsubsection{Consequences for the model and predictions}

Taking this excess seriously has interesting consequences for the model parameters.
First, a charged Higgs with a mass of a few hundred GeV is predicted in any case.
Second,
the $\nu_R$ of muon type must be light, 
\begin{align}
m_{\nu_{R,\mu}} \lsim 200~{\rm GeV}\,.
\end{align}
The third right-handed neutrino, in turn, is only constrained by the requirement that 
$H^{\pm \pm}$ is heavy enough and the vacuum is stable.
In most scenarios we find the $H^{\pm\pm}$ mass to be light enough to be soon detected at the LHC. 
As $y_4^L$ is responsible for the $\nu_R$ mass as well as the $H^{\pm\pm}$ coupling to charged leptons, 
the current bounds on $m_{H^{\pm \pm}}$ are already at $\simeq 440~$GeV if $\nu_{R,e}$ 
is the heaviest $\nu_R$ eigenstate. 
$H^{\pm\pm}$ masses beyond 500~GeV are hard to achieve; see the discussion in Ref.~\cite{Basso:2015pka}.

The last consequence is, of course, that a $W'$ has to be measured in the near future.
If the $W'$ in fact gets measured with $M_{W'} \approx 2~$TeV, then 
the cross sections necessary for the $eejj$ measurement imply that our model predicts 
additional sub-TeV supersymmetric particles that the $W'$ and the $\nu_R$ can decay to.
If, in turn, it is measured at 2.3~TeV, this implies
that 
the supersymmetric spectrum cannot be that light and that  in particular the Higgsino states 
have masses above  $M_{W'}/2$.

\section{Conclusions}

An excess of events with two electrons and two jets has been observed with the CMS detector
with a local significance of 2.8~$\sigma$. While the simplest left-right symmetric models 
fail to explain this excess, several realizations have been proposed in which a 
$W'$ and a right-handed neutrino could be the origin of this signal.
In this paper, we have stressed that in a complete model both the $W'$ as well as the $\nu_R$
have in general additional decay modes. Taking them into account reduces the signal strength
of the $eejj$ final state.
In our particular supersymmetric realization, a light charged Higgs boson opens
the possibility of the two-body decay $\nu_R \to \ell^\pm H^\mp_1$. Once this is taken
into account, we can explain the observed excess. Interestingly, the data can be
accommodated better if light Higgsino-like charginos/neutralinos are present, to which
the $W'$ can decay. This in combination with the two-body decay implies that $M_{W'}$ should be
about 2 TeV. If the SUSY spectrum is heavier, then a slightly larger $M_{W'} \simeq 2.3$~TeV
 is preferred. In both cases, this model predicts a relatively light  $\nu_R$ of muon type
 with a mass below  200 GeV, and also a doubly charged Higgs boson should be detected soon
at the LHC.

\section*{Acknowledgement}

We thank Lorenzo Basso and Benjamin Fuks for interesting discussions which led to this study. 
This work has been supported by  the DFG, Project No.\ PO-1337/3-1.


\begin{thebibliography}{10}%
\makeatletter
\providecommand \@ifxundefined [1]{%
 \ifx #1\undefined \expandafter \@firstoftwo
 \else \expandafter \@secondoftwo
\fi
}%
\providecommand \@ifnum [1]{%
 \ifnum #1\expandafter \@firstoftwo
 \else \expandafter \@secondoftwo
\fi
}%
\providecommand \enquote [1]{``#1''}%
\providecommand \bibnamefont  [1]{#1}%
\providecommand \bibfnamefont [1]{#1}%
\providecommand \citenamefont [1]{#1}%
\providecommand\href[0]{\@sanitize\@href}%
\providecommand\@href[1]{\endgroup\@@startlink{#1}\endgroup\@@href}%
\providecommand\@@href[1]{#1\@@endlink}%
\providecommand \@sanitize [0]{\begingroup\catcode`\&12\catcode`\#12\relax}%
\@ifxundefined \pdfoutput {\@firstoftwo}{%
 \@ifnum{\z@=\pdfoutput}{\@firstoftwo}{\@secondoftwo}%
}{%
 \providecommand\@@startlink[1]{\leavevmode}%
 \providecommand\@@endlink[0]{}%
}{%
 \providecommand\@@startlink[1]{%
  \leavevmode
  \pdfstartlink
   attr{/Border[0 0 1 ]/H/I/C[0 1 1]}%
   user{/Subtype/Link/A<</Type/Action/S/URI/URI(#1)>>}%
  \relax
 }%
 \providecommand\@@endlink[0]{\pdfendlink}%
}%
\providecommand \url  [0]{\begingroup\@sanitize \@url }%
\providecommand \@url [1]{\endgroup\@href {#1}{\urlprefix}}%
\providecommand \urlprefix [0]{URL }%
\providecommand \Eprint[0]{\href }%
\@ifxundefined \urlstyle {%
  \providecommand \doi [1]{doi:\discretionary{}{}{}#1}%
}{%
  \providecommand \doi [0]{doi:\discretionary{}{}{}\begingroup
  \urlstyle{rm}\Url }%
}%
\providecommand \doibase [0]{http://dx.doi.org/}%
\providecommand \Doi[1]{\href{\doibase#1}}%
\providecommand \bibAnnote [3]{%
  \BibitemShut{#1}%
  \begin{quotation}\noindent
    \textsc{Key:}\ #2\\\textsc{Annotation:}\ #3%
  \end{quotation}%
}%
\providecommand \bibAnnoteFile [2]{%
  \IfFileExists{#2}{\bibAnnote {#1} {#2} {\input{#2}}}{}%
}%
\providecommand \typeout [0]{\immediate \write \m@ne }%
\providecommand \selectlanguage [0]{\@gobble}%
\providecommand \bibinfo [0]{\@secondoftwo}%
\providecommand \bibfield [0]{\@secondoftwo}%
\providecommand \translation [1]{[#1]}%
\providecommand \BibitemOpen[0]{}%
\providecommand \bibitemStop [0]{}%
\providecommand \bibitemNoStop [0]{.\EOS\space}%
\providecommand \EOS [0]{\spacefactor3000\relax}%
\providecommand \BibitemShut [1]{\csname bibitem#1\endcsname}%
\bibitem{ATLAS:2012ae}%
  \BibitemOpen
  \bibfield{author}{%
  \bibinfo {author} {\bibfnamefont{G.}~\bibnamefont{Aad}} \emph{et~al.}
  (\bibinfo {collaboration} {ATLAS}),\ }%
  \bibfield{journal}{%
  \Doi{10.1016/j.physletb.2012.02.044}{\bibinfo {journal} {Phys.Lett.}}\ }%
  \textbf{\bibinfo {volume} {B710}},\ \bibinfo {pages} {49} (\bibinfo {year}
  {2012}),\ \Eprint{http://arxiv.org/abs/1202.1408}{arXiv:1202.1408 [hep-ex]}%
  \bibAnnoteFile{NoStop}{ATLAS:2012ae}%
\bibitem{Chatrchyan:2012tx}%
  \BibitemOpen
  \bibfield{author}{%
  \bibinfo {author} {\bibfnamefont{S.}~\bibnamefont{Chatrchyan}} \emph{et~al.}
  (\bibinfo {collaboration} {CMS}),\ }%
  \bibfield{journal}{%
  \Doi{10.1016/j.physletb.2012.02.064}{\bibinfo {journal} {Phys.Lett.}}\ }%
  \textbf{\bibinfo {volume} {B710}},\ \bibinfo {pages} {26} (\bibinfo {year}
  {2012}),\ \Eprint{http://arxiv.org/abs/1202.1488}{arXiv:1202.1488 [hep-ex]}%
  \bibAnnoteFile{NoStop}{Chatrchyan:2012tx}%
\bibitem{Bechtle:2015nua}%
  \BibitemOpen
  \bibfield{author}{%
  \bibinfo {author} {\bibfnamefont{P.}~\bibnamefont{Bechtle}}, \bibinfo {author} {\bibfnamefont{J.~E.}~\bibnamefont{Camargo-Molina}}, \bibinfo {author} {\bibfnamefont{K.}~\bibnamefont{Desch}}, \bibinfo {author} {\bibfnamefont{H.}~\bibnamefont{Dreiner}}, \bibinfo {author} {\bibfnamefont{M.}~\bibnamefont{Hamer}},  \emph{et~al.},\ }%
  (\bibinfo {year}  {2015}),\
  \Eprint{http://arxiv.org/abs/1508.05951}{arXiv:1508.05951 [hep-ph]}%
  \bibAnnoteFile{NoStop}{Chatrchyan:2012tx}%
\bibitem{Babu:1987kp}%
  \BibitemOpen
  \bibfield{author}{%
  \bibinfo {author} {\bibfnamefont{K.~S.}~\bibnamefont{Babu}}, \bibinfo {author}
  {\bibfnamefont{X.-G.}\ \bibnamefont{He}},\ and\ \bibinfo {author}
  {\bibfnamefont{E.}~\bibnamefont{Ma}},\ }%
  \bibfield{journal}{%
  \Doi{10.1103/PhysRevD.36.878}{\bibinfo {journal} {Phys.Rev.}}\ }%
  \textbf{\bibinfo {volume} {D36}},\ \bibinfo {pages} {878} (\bibinfo {year}
  {1987})%
  \bibAnnoteFile{NoStop}{Babu:1987kp}%
\bibitem{Zhang:2008jm}%
  \BibitemOpen
  \bibfield{author}{%
  \bibinfo {author} {\bibfnamefont{Y.}~\bibnamefont{Zhang}}, \bibinfo {author}
  {\bibfnamefont{H.}~\bibnamefont{An}}, \bibinfo {author}
  {\bibfnamefont{X.-d.}\ \bibnamefont{Ji}},\ and\ \bibinfo {author}
  {\bibfnamefont{R.~N.}\ \bibnamefont{Mohapatra}},\ }%
  \bibfield{journal}{%
  \Doi{10.1103/PhysRevD.78.011302}{\bibinfo {journal} {Phys.Rev.}}\ }%
  \textbf{\bibinfo {volume} {D78}},\ \bibinfo {pages} {011302} (\bibinfo {year}
  {2008}),\ \Eprint{http://arxiv.org/abs/0804.0268}{arXiv:0804.0268 [hep-ph]}%
  \bibAnnoteFile{NoStop}{Zhang:2008jm}%
\bibitem{Hirsch:2011hg}%
  \BibitemOpen
  \bibfield{author}{%
  \bibinfo {author} {\bibfnamefont{M.}~\bibnamefont{Hirsch}}, \bibinfo {author}
  {\bibfnamefont{M.}~\bibnamefont{Malinsky}}, \bibinfo {author}
  {\bibfnamefont{W.}~\bibnamefont{Porod}}, \bibinfo {author}
  {\bibfnamefont{L.}~\bibnamefont{Reichert}},\ and\ \bibinfo {author}
  {\bibfnamefont{F.}~\bibnamefont{Staub}},\ }%
  \bibfield{journal}{%
  \Doi{10.1007/JHEP02(2012)084}{\bibinfo {journal} {JHEP}}\ }%
  \textbf{\bibinfo {volume} {02}},\ \bibinfo {pages} {084} (\bibinfo {year}
  {2012}),\ \Eprint{http://arxiv.org/abs/1110.3037}{arXiv:1110.3037 [hep-ph]}%
  \bibAnnoteFile{NoStop}{Hirsch:2011hg}%
\bibitem{Krauss:2013jva}%
  \BibitemOpen
  \bibfield{author}{%
  \bibinfo {author} {\bibfnamefont{M.~E.}\ \bibnamefont{Krauss}}, \bibinfo
  {author} {\bibfnamefont{W.}~\bibnamefont{Porod}},\ and\ \bibinfo {author}
  {\bibfnamefont{F.}~\bibnamefont{Staub}},\ }%
  \bibfield{journal}{%
  \Doi{10.1103/PhysRevD.88.015014}{\bibinfo {journal} {Phys.Rev.}}\ }%
  \textbf{\bibinfo {volume} {D88}},\ \bibinfo {pages} {015014} (\bibinfo {year}
  {2013}),\ \Eprint{http://arxiv.org/abs/1304.0769}{arXiv:1304.0769 [hep-ph]}%
  \bibAnnoteFile{NoStop}{Krauss:2013jva}%
\bibitem{Khachatryan:2014dka}%
  \BibitemOpen
  \bibfield{author}{%
  \bibinfo {author} {\bibfnamefont{V.}~\bibnamefont{Khachatryan}} \emph{et~al.}
  (\bibinfo {collaboration} {CMS}),\ }%
  \bibfield{journal}{%
  \Doi{10.1140/epjc/s10052-014-3149-z}{\bibinfo {journal} {Eur.Phys.J.}}\ }%
  \textbf{\bibinfo {volume} {C74}},\ \bibinfo {pages} {3149} (\bibinfo {year}
  {2014}),\ \Eprint{http://arxiv.org/abs/1407.3683}{arXiv:1407.3683 [hep-ex]}%
  \bibAnnoteFile{NoStop}{Khachatryan:2014dka}%
\bibitem{Deppisch:2014qpa}%
  \BibitemOpen
  \bibfield{author}{%
  \bibinfo {author} {\bibfnamefont{F.~F.}\ \bibnamefont{Deppisch}}, \bibinfo
  {author} {\bibfnamefont{T.~E.}\ \bibnamefont{Gonzalo}}, \bibinfo {author}
  {\bibfnamefont{S.}~\bibnamefont{Patra}}, \bibinfo {author}
  {\bibfnamefont{N.}~\bibnamefont{Sahu}},\ and\ \bibinfo {author}
  {\bibfnamefont{U.}~\bibnamefont{Sarkar}},\ }%
  \bibfield{journal}{%
  \Doi{10.1103/PhysRevD.90.053014}{\bibinfo {journal} {Phys.Rev.}}\ }%
  \textbf{\bibinfo {volume} {D90}},\ \bibinfo {pages} {053014} (\bibinfo {year}
  {2014}),\ \Eprint{http://arxiv.org/abs/1407.5384}{arXiv:1407.5384 [hep-ph]}%
  \bibAnnoteFile{NoStop}{Deppisch:2014qpa}%
\bibitem{Heikinheimo:2014tba}%
  \BibitemOpen
  \bibfield{author}{%
  \bibinfo {author} {\bibfnamefont{M.}\ \bibnamefont{Heikinheimo}}, \bibinfo
  {author} {\bibfnamefont{M.}\ \bibnamefont{Raidal}}, \bibinfo {author}
  {\bibfnamefont{C.}~\bibnamefont{Spethmann}},\ }%
  \bibfield{journal}{%
  \Doi{10.1140/epjc/s10052-014-3107-9}{\bibinfo {journal} {Eur.Phys.J.}}\ }%
  \textbf{\bibinfo {volume} {C74}},\ \bibinfo {pages} {3107} (\bibinfo {year}
  {2014}),\ \Eprint{http://arxiv.org/abs/1407.6908}{arXiv:1407.6908 [hep-ph]}%
  \bibAnnoteFile{NoStop}{Heikinheimo:2014tba}%
\bibitem{Aguilar-Saavedra:2014ola}%
  \BibitemOpen
  \bibfield{author}{%
  \bibinfo {author} {\bibfnamefont{J.~A.}~\bibnamefont{Aguilar-Saavedra}}\ and\
  \bibinfo {author} {\bibfnamefont{F.~R.}~\bibnamefont{Joaquim}},\ }%
  \bibfield{journal}{%
  \Doi{10.1103/PhysRevD.90.115010}{\bibinfo {journal} {Phys.Rev.}}\ }%
  \textbf{\bibinfo {volume} {D90}},\ \bibinfo {pages} {115010} (\bibinfo {year}
  {2014}),\ \Eprint{http://arxiv.org/abs/1408.2456}{arXiv:1408.2456 [hep-ph]}%
  \bibAnnoteFile{NoStop}{Aguilar-Saavedra:2014ola}%
\bibitem{Gluza:2015goa}%
  \BibitemOpen
  \bibfield{author}{%
  \bibinfo {author} {\bibfnamefont{J.}~\bibnamefont{Gluza}}\ and\ \bibinfo
  {author} {\bibfnamefont{T.}~\bibnamefont{Jelinski}},\ }%
  \bibfield{journal}{%
  \Doi{http://dx.doi.org/10.1016/j.physletb.2015.06.077}{\bibinfo {journal}
  {Phys.Lett.}}\ }%
  \textbf{\bibinfo {volume} {B748}},\ \bibinfo {pages} {125 } (\bibinfo {year}
  {2015}),\ \Eprint{http://arxiv.org/abs/1504.05568}{arXiv:1504.05568}%
  \bibAnnoteFile{NoStop}{Gluza:2015goa}%
\bibitem{Allanach:2014nna}%
  \BibitemOpen
  \bibfield{author}{%
  \bibinfo {author} {\bibfnamefont{B.}~\bibnamefont{Allanach}}, \bibinfo
  {author} {\bibfnamefont{S.}~\bibnamefont{Biswas}}, \bibinfo {author}
  {\bibfnamefont{S.}~\bibnamefont{Mondal}},\ and\ \bibinfo {author}
  {\bibfnamefont{M.}~\bibnamefont{Mitra}},\ }%
  \bibfield{journal}{%
  \Doi{10.1103/PhysRevD.91.015011}{\bibinfo {journal} {Phys.Rev.}}\ }%
  \textbf{\bibinfo {volume} {D91}},\ \bibinfo {pages} {015011} (\bibinfo {year}
  {2015}),\ \Eprint{http://arxiv.org/abs/1410.5947}{arXiv:1410.5947 [hep-ph]}%
  \bibAnnoteFile{NoStop}{Allanach:2014nna}%
\bibitem{Allanach:2014lca}%
  \BibitemOpen
  \bibfield{author}{%
  \bibinfo {author} {\bibfnamefont{B.}~\bibnamefont{Allanach}}, \bibinfo
  {author} {\bibfnamefont{S.}~\bibnamefont{Biswas}}, \bibinfo {author}
  {\bibfnamefont{S.}~\bibnamefont{Mondal}},\ and\ \bibinfo {author}
  {\bibfnamefont{M.}~\bibnamefont{Mitra}},\ }%
  \bibfield{journal}{%
  \Doi{10.1103/PhysRevD.91.011702}{\bibinfo {journal} {Phys.Rev.}}\ }%
  \textbf{\bibinfo {volume} {D91}},\ \bibinfo {pages} {011702} (\bibinfo {year}
  {2015}),\ \Eprint{http://arxiv.org/abs/1408.5439}{arXiv:1408.5439 [hep-ph]}%
  \bibAnnoteFile{NoStop}{Allanach:2014lca}%
\bibitem{Dobrescu:2014esa}%
  \BibitemOpen
  \bibfield{author}{%
  \bibinfo {author} {\bibfnamefont{B.~A.}\ \bibnamefont{Dobrescu}}\ and\
  \bibinfo {author} {\bibfnamefont{A.}~\bibnamefont{Martin}},\ }%
  \bibfield{journal}{%
  \Doi{10.1103/PhysRevD.91.035019}{\bibinfo {journal} {Phys.Rev.}}\ }%
  \textbf{\bibinfo {volume} {D91}},\ \bibinfo {pages} {035019} (\bibinfo {year}
  {2015}),\ \Eprint{http://arxiv.org/abs/1408.1082}{arXiv:1408.1082 [hep-ph]}%
  \bibAnnoteFile{NoStop}{Dobrescu:2014esa}%
\bibitem{Dhuria:2015wwa}%
  \BibitemOpen
  \bibfield{author}{%
  \bibinfo {author} {\bibfnamefont{M.}~\bibnamefont{Dhuria}}, \bibinfo {author}
  {\bibfnamefont{C.}~\bibnamefont{Hati}}, \bibinfo {author}
  {\bibfnamefont{R.}~\bibnamefont{Rangarajan}},\ and\ \bibinfo {author}
  {\bibfnamefont{U.}~\bibnamefont{Sarkar}}}\ %
   (\bibinfo {year} {2015}),\
  \Eprint{http://arxiv.org/abs/1502.01695}{arXiv:1502.01695 [hep-ph]}%
  \bibAnnoteFile{NoStop}{Dhuria:2015wwa}%
\bibitem{Dhuria:2015hta}%
  \BibitemOpen
  \bibfield{author}{%
  \bibinfo {author} {\bibfnamefont{M.}~\bibnamefont{Dhuria}}, \bibinfo {author}
  {\bibfnamefont{C.}~\bibnamefont{Hati}}, \bibinfo {author}
  {\bibfnamefont{R.}~\bibnamefont{Rangarajan}},\ and\ \bibinfo {author}
  {\bibfnamefont{U.}~\bibnamefont{Sarkar}},\ }%
  \bibfield{journal}{%
  \Doi{10.1103/PhysRevD.91.055010}{\bibinfo {journal} {Phys.Rev.}}\ }%
  \textbf{\bibinfo {volume} {D91}},\ \bibinfo {pages} {055010} (\bibinfo {year}
  {2015}),\ \Eprint{http://arxiv.org/abs/1501.04815}{arXiv:1501.04815
  [hep-ph]}%
  \bibAnnoteFile{NoStop}{Dhuria:2015hta}%
\bibitem{Brooijmans:2014eja}%
  \BibitemOpen
  \bibfield{author}{%
  \bibinfo {author} {\bibfnamefont{G.}~\bibnamefont{Brooijmans}}, \bibinfo
  {author} {\bibfnamefont{R.}~\bibnamefont{Contino}}, \bibinfo {author}
  {\bibfnamefont{B.}~\bibnamefont{Fuks}}, \bibinfo {author}
  {\bibfnamefont{F.}~\bibnamefont{Moortgat}}, \bibinfo {author}
  {\bibfnamefont{P.}~\bibnamefont{Richardson}}, \emph{et~al.}}\ %
   (\bibinfo {year} {2014}),\
  \Eprint{http://arxiv.org/abs/1405.1617}{arXiv:1405.1617 [hep-ph]}%
  \bibAnnoteFile{NoStop}{Brooijmans:2014eja}%
\bibitem{Basso:2015pka}%
  \BibitemOpen
  \bibfield{author}{%
  \bibinfo {author} {\bibfnamefont{L.}~\bibnamefont{Basso}}, \bibinfo {author}
  {\bibfnamefont{B.}~\bibnamefont{Fuks}}, \bibinfo {author}
  {\bibfnamefont{M.~E.}\ \bibnamefont{Krauss}},\ and\ \bibinfo {author}
  {\bibfnamefont{W.}~\bibnamefont{Porod}}}\ %
  \bibfield{journal}{%
  \Doi{10.1007/JHEP07(2015)147}{\bibinfo {journal} {JHEP}}\ }%
  \textbf{\bibinfo {volume} {1507}},\ \bibinfo {pages} {147} (\bibinfo {year} {2015}),\
  \Eprint{http://arxiv.org/abs/1503.08211}{arXiv:1503.08211 [hep-ph]}%
  \bibAnnoteFile{NoStop}{Basso:2015pka}%
\bibitem{Alloul:2013fra}%
  \BibitemOpen
  \bibfield{author}{%
  \bibinfo {author} {\bibfnamefont{A.}~\bibnamefont{Alloul}}, \bibinfo {author}
  {\bibfnamefont{M.}~\bibnamefont{Frank}}, \bibinfo {author}
  {\bibfnamefont{B.}~\bibnamefont{Fuks}},\ and\ \bibinfo {author}
  {\bibfnamefont{M.}~\bibnamefont{Rausch~de Traubenberg}},\ }%
  \bibfield{journal}{%
  \Doi{10.1007/JHEP10(2013)033}{\bibinfo {journal} {JHEP}}\ }%
  \textbf{\bibinfo {volume} {1310}},\ \bibinfo {pages} {033} (\bibinfo {year}
  {2013}),\ \Eprint{http://arxiv.org/abs/1307.5073}{arXiv:1307.5073 [hep-ph]}%
  \bibAnnoteFile{NoStop}{Alloul:2013fra}%
\bibitem{Kuchimanchi:1993jg}%
  \BibitemOpen
  \bibfield{author}{%
  \bibinfo {author} {\bibfnamefont{R.}~\bibnamefont{Kuchimanchi}}\ and\
  \bibinfo {author} {\bibfnamefont{R.~N.}~\bibnamefont{Mohapatra}},\ }%
  \bibfield{journal}{%
  \Doi{10.1103/PhysRevD.48.4352}{\bibinfo {journal} {Phys.Rev.}}\ }%
  \textbf{\bibinfo {volume} {D48}},\ \bibinfo {pages} {4352} (\bibinfo {year}
  {1993}),\ \Eprint{http://arxiv.org/abs/hep-ph/9306290}{arXiv:hep-ph/9306290}%
  \bibAnnoteFile{NoStop}{Kuchimanchi:1993jg}%
\bibitem{Huitu:1994zm}%
  \BibitemOpen
  \bibfield{author}{%
  \bibinfo {author} {\bibfnamefont{K.}~\bibnamefont{Huitu}}\ and\ \bibinfo
  {author} {\bibfnamefont{J.}~\bibnamefont{Maalampi}},\ }%
  \bibfield{journal}{%
  \Doi{10.1016/0370-2693(94)01531-G}{\bibinfo {journal} {Phys.Lett.}}\ }%
  \textbf{\bibinfo {volume} {B344}},\ \bibinfo {pages} {217} (\bibinfo {year}
  {1995}),\ \Eprint{http://arxiv.org/abs/hep-ph/9410342}{arXiv:hep-ph/9410342}%
  \bibAnnoteFile{NoStop}{Huitu:1994zm}%
\bibitem{Babu:2008ep}%
  \BibitemOpen
  \bibfield{author}{%
  \bibinfo {author} {\bibfnamefont{K.}~\bibnamefont{Babu}}\ and\ \bibinfo
  {author} {\bibfnamefont{R.~N.}\ \bibnamefont{Mohapatra}},\ }%
  \bibfield{journal}{%
  \Doi{10.1016/j.physletb.2008.09.018}{\bibinfo {journal} {Phys.Lett.}}\ }%
  \textbf{\bibinfo {volume} {B668}},\ \bibinfo {pages} {404} (\bibinfo {year}
  {2008}),\ \Eprint{http://arxiv.org/abs/0807.0481}{arXiv:0807.0481 [hep-ph]}%
  \bibAnnoteFile{NoStop}{Babu:2008ep}%
\bibitem{Babu:2014vba}%
  \BibitemOpen
  \bibfield{author}{%
  \bibinfo {author} {\bibfnamefont{K.}~\bibnamefont{Babu}}\ and\ \bibinfo
  {author} {\bibfnamefont{A.}~\bibnamefont{Patra}}}\ %
   (\bibinfo {year} {2014}),\
  \Eprint{http://arxiv.org/abs/1412.8714}{arXiv:1412.8714 [hep-ph]}%
  \bibAnnoteFile{NoStop}{Babu:2014vba}%
\bibitem{Brehmer:2015cia}%
  \BibitemOpen
  \bibfield{author}{%
  \bibinfo {author} {\bibfnamefont{J.}~\bibnamefont{Brehmer}}, \bibinfo
  {author} {\bibfnamefont{J.}~\bibnamefont{Hewett}}, \bibinfo {author}
  {\bibfnamefont{J.}~\bibnamefont{Kopp}}, \bibinfo {author}
  {\bibfnamefont{T.}~\bibnamefont{Rizzo}},\ and\ \bibinfo {author}
  {\bibfnamefont{J.}~\bibnamefont{Tattersall}}}\ %
   (\bibinfo {year} {2015}),\
  \Eprint{http://arxiv.org/abs/1507.00013}{arXiv:1507.00013 [hep-ph]}%
  \bibAnnoteFile{NoStop}{Brehmer:2015cia}%
\bibitem{Dobrescu:2015yba}%
  \BibitemOpen
  \bibfield{author}{%
  \bibinfo {author} {\bibfnamefont{B.~A.}\ \bibnamefont{Dobrescu}}\ and\
  \bibinfo {author} {\bibfnamefont{Z.}~\bibnamefont{Liu}}}\ %
   (\bibinfo {year} {2015}),\
  \Eprint{http://arxiv.org/abs/1507.01923}{arXiv:1507.01923 [hep-ph]}%
  \bibAnnoteFile{NoStop}{Dobrescu:2015yba}%
\bibitem{Aad:2015owa}%
  \BibitemOpen
  \bibfield{author}{%
  \bibinfo {author} {\bibfnamefont{G.}~\bibnamefont{Aad}} \emph{et~al.}
  (\bibinfo {collaboration} {ATLAS})}\ %
   (\bibinfo {year} {2015}),\
  \Eprint{http://arxiv.org/abs/1506.00962}{arXiv:1506.00962 [hep-ex]}%
  \bibAnnoteFile{NoStop}{Aad:2015owa}%
\bibitem{Aad:2015xaa}%
  \BibitemOpen
  \bibfield{author}{%
  \bibinfo {author} {\bibfnamefont{G.}~\bibnamefont{Aad}} \emph{et~al.}
  (\bibinfo {collaboration} {ATLAS})}\ %
   (\bibinfo {year} {2015}),\
  \Eprint{http://arxiv.org/abs/1506.06020}{arXiv:1506.06020 [hep-ex]}%
  \bibAnnoteFile{NoStop}{Aad:2015xaa}%
\bibitem{Chatrchyan:2014koa}%
  \BibitemOpen
  \bibfield{author}{%
  \bibinfo {author} {\bibfnamefont{S.}~\bibnamefont{Chatrchyan}} \emph{et~al.}
  (\bibinfo {collaboration} {CMS}),\ }%
  \bibfield{journal}{%
  \Doi{10.1007/JHEP05(2014)108}{\bibinfo {journal} {JHEP}}\ }%
  \textbf{\bibinfo {volume} {1405}},\ \bibinfo {pages} {108} (\bibinfo {year}
  {2014}),\ \Eprint{http://arxiv.org/abs/1402.2176}{arXiv:1402.2176 [hep-ex]}%
  \bibAnnoteFile{NoStop}{Chatrchyan:2014koa}%
\bibitem{Khachatryan:2015sja}%
  \BibitemOpen
  \bibfield{author}{%
  \bibinfo {author} {\bibfnamefont{V.}~\bibnamefont{Khachatryan}} \emph{et~al.}
  (\bibinfo {collaboration} {CMS}),\ }%
  \bibfield{journal}{%
  \Doi{10.1103/PhysRevD.91.052009}{\bibinfo {journal} {Phys.Rev.}}\ }%
  \textbf{\bibinfo {volume} {D91}},\ \bibinfo {pages} {052009} (\bibinfo {year}
  {2015}),\ \Eprint{http://arxiv.org/abs/1501.04198}{arXiv:1501.04198
  [hep-ex]}%
  \bibAnnoteFile{NoStop}{Khachatryan:2015sja}%
\bibitem{Bertolini:2014sua}
  \BibitemOpen
  \bibfield{author}{%
  \bibinfo {author} {\bibfnamefont{S.}~\bibnamefont{Bertolini}}, \bibinfo
  {author} {\bibfnamefont{A.}~\bibnamefont{Maiezza}}, \bibinfo {author}
  {\bibfnamefont{F.}~\bibnamefont{Nesti}},\ }%
  \bibfield{journal}{%
  \Doi{10.1103/PhysRevD.89.095028}{\bibinfo {journal} {Phys.Rev.}}\ }%
  \textbf{\bibinfo {volume} {D89}},\ \bibinfo {pages} {095028} (\bibinfo {year}
  {2014}),\ \Eprint{http://arxiv.org/abs/1403.7112}{arXiv:1403.7112 [hep-ph]}%
  \bibAnnoteFile{NoStop}{Bertolini:2014sua}%
\bibitem{Maiezza:2010ic}
  \BibitemOpen
  \bibfield{author}{%
  \bibinfo {author} {\bibfnamefont{A.}~\bibnamefont{Maiezza}}, \bibinfo
  {author} {\bibfnamefont{M.}~\bibnamefont{Nemevsek}}, \bibinfo {author}
  {\bibfnamefont{F.}~\bibnamefont{Nesti}}, \bibinfo {author}
  {\bibfnamefont{G.}~\bibnamefont{Senjanovic}}\ }%
  \bibfield{journal}{%
  \Doi{10.1103/PhysRevD.82.055022}{\bibinfo {journal} {Phys.Rev.}}\ }%
  \textbf{\bibinfo {volume} {D82}},\ \bibinfo {pages} {055022} (\bibinfo {year}
  {2010}),\ \Eprint{http://arxiv.org/abs/1005.5160}{arXiv:1005.5160 [hep-ph]}%
  \bibAnnoteFile{NoStop}{Bertolini:2014sua}%
\bibitem{Guadagnoli:2010sd}
  \BibitemOpen
  \bibfield{author}{%
  \bibinfo {author} {\bibfnamefont{D.}~\bibnamefont{Guadagnoli}}, \bibinfo
  {author} {\bibfnamefont{R.~N.}~\bibnamefont{Mohapatra}},\ }%
  \bibfield{journal}{%
  \Doi{10.1016/j.physletb.2010.10.027}{\bibinfo {journal} {Phys.Lett.}}\ }%
  \textbf{\bibinfo {volume} {B694}},\ \bibinfo {pages} {386} (\bibinfo {year}
  {2011}),\ \Eprint{http://arxiv.org/abs/1008.1074}{arXiv:1008.1074 [hep-ph]}%
  \bibAnnoteFile{NoStop}{Guadagnoli:2010sd}%
\bibitem{Zhang:2007qma}
  \BibitemOpen
  \bibfield{author}{%
  \bibinfo {author} {\bibfnamefont{Y.}~\bibnamefont{Zhang}}, \bibinfo
  {author} {\bibfnamefont{H.}~\bibnamefont{An}}, \bibinfo {author}
  {\bibfnamefont{X.}~\bibnamefont{Ji}},\ }%
  \bibfield{journal}{%
  \Doi{10.1103/PhysRevD.78.035006}{\bibinfo {journal} {Phys.Rev.}}\ }%
  \textbf{\bibinfo {volume} {D78}},\ \bibinfo {pages} {035006} (\bibinfo {year}
  {2008}),\ \Eprint{http://arxiv.org/abs/0710.1454}{arXiv:0710.1454 [hep-ph]}%
  \bibAnnoteFile{NoStop}{Bertolini:2014sua}%
  %
  \bibitem{Chatrchyan:2012ya}%
  \BibitemOpen
  \bibfield{author}{%
  \bibinfo {author} {\bibfnamefont{S.}~\bibnamefont{Chatrchyan}} \emph{et~al.}
  (\bibinfo {collaboration} {CMS Collaboration}),\ }%
  \bibfield{journal}{%
  \Doi{10.1140/epjc/s10052-012-2189-5}{\bibinfo {journal} {Eur.Phys.J.}}\ }%
  \textbf{\bibinfo {volume} {C72}},\ \bibinfo {pages} {2189} (\bibinfo {year}
  {2012}),\ \Eprint{http://arxiv.org/abs/1207.2666}{arXiv:1207.2666 [hep-ex]}%
  \bibAnnoteFile{NoStop}{Chatrchyan:2012ya}%
\bibitem{BarShalom:2008gt}%
  \BibitemOpen
  \bibfield{author}{%
  \bibinfo {author} {\bibfnamefont{S.}~\bibnamefont{Bar-Shalom}}, \bibinfo
  {author} {\bibfnamefont{G.}~\bibnamefont{Eilam}}, \bibinfo {author}
  {\bibfnamefont{T.}~\bibnamefont{Han}},\ and\ \bibinfo {author}
  {\bibfnamefont{A.}~\bibnamefont{Soni}},\ }%
  \bibfield{journal}{%
  \Doi{10.1103/PhysRevD.77.115019}{\bibinfo {journal} {Phys.Rev.}}\ }%
  \textbf{\bibinfo {volume} {D77}},\ \bibinfo {pages} {115019} (\bibinfo {year}
  {2008}),\ \Eprint{http://arxiv.org/abs/0803.2835}{arXiv:0803.2835 [hep-ph]}%
  \bibAnnoteFile{NoStop}{BarShalom:2008gt}%
\bibitem{Staub:2008uz}%
  \BibitemOpen
  \bibfield{author}{%
  \bibinfo {author} {\bibfnamefont{F.}~\bibnamefont{Staub}}}\ %
   (\bibinfo {year} {2008}),\
  \Eprint{http://arxiv.org/abs/0806.0538}{arXiv:0806.0538 [hep-ph]}%
  \bibAnnoteFile{NoStop}{Staub:2008uz}%
\bibitem{Staub:2009bi}%
  \BibitemOpen
  \bibfield{author}{%
  \bibinfo {author} {\bibfnamefont{F.}~\bibnamefont{Staub}},\ }%
  \bibfield{journal}{%
  \Doi{10.1016/j.cpc.2010.01.011}{\bibinfo {journal} {Comput.Phys.Commun.}}\ }%
  \textbf{\bibinfo {volume} {181}},\ \bibinfo {pages} {1077} (\bibinfo {year}
  {2010}),\ \Eprint{http://arxiv.org/abs/0909.2863}{arXiv:0909.2863 [hep-ph]}%
  \bibAnnoteFile{NoStop}{Staub:2009bi}%
\bibitem{Staub:2010jh}%
  \BibitemOpen
  \bibfield{author}{%
  \bibinfo {author} {\bibfnamefont{F.}~\bibnamefont{Staub}},\ }%
  \bibfield{journal}{%
  \Doi{10.1016/j.cpc.2010.11.030}{\bibinfo {journal} {Comput.Phys.Commun.}}\ }%
  \textbf{\bibinfo {volume} {182}},\ \bibinfo {pages} {808} (\bibinfo {year}
  {2011}),\ \Eprint{http://arxiv.org/abs/1002.0840}{arXiv:1002.0840 [hep-ph]}%
  \bibAnnoteFile{NoStop}{Staub:2010jh}%
\bibitem{Staub:2012pb}%
  \BibitemOpen
  \bibfield{author}{%
  \bibinfo {author} {\bibfnamefont{F.}~\bibnamefont{Staub}},\ }%
  \bibfield{journal}{%
  \Doi{10.1016/j.cpc.2013.02.019}{\bibinfo {journal} {Computer Physics
  Communications}}\ }%
  \textbf{\bibinfo {volume} {184}},\ \bibinfo {pages} {pp. 1792} (\bibinfo
  {year} {2013}),\ \Eprint{http://arxiv.org/abs/1207.0906}{arXiv:1207.0906
  [hep-ph]}%
  \bibAnnoteFile{NoStop}{Staub:2012pb}%
\bibitem{Staub:2013tta}%
  \BibitemOpen
  \bibfield{author}{%
  \bibinfo {author} {\bibfnamefont{F.}~\bibnamefont{Staub}},\ }%
  \bibfield{journal}{%
  \Doi{10.1016/j.cpc.2014.02.018}{\bibinfo {journal} {Comput. Phys. Commun.}}\
  }%
  \textbf{\bibinfo {volume} {185}},\ \bibinfo {pages} {1773} (\bibinfo {year}
  {2014}),\ \Eprint{http://arxiv.org/abs/1309.7223}{arXiv:1309.7223 [hep-ph]}%
  \bibAnnoteFile{NoStop}{Staub:2013tta}%
\bibitem{Staub:2015kfa}%
  \BibitemOpen
  \bibfield{author}{%
  \bibinfo {author} {\bibfnamefont{F.}~\bibnamefont{Staub}}}\ %
   (\bibinfo {year} {2015}),\
  \Eprint{http://arxiv.org/abs/1503.04200}{arXiv:1503.04200 [hep-ph]}%
  \bibAnnoteFile{NoStop}{Staub:2015kfa}%
\bibitem{Porod:2003um}%
  \BibitemOpen
  \bibfield{author}{%
  \bibinfo {author} {\bibfnamefont{W.}~\bibnamefont{Porod}},\ }%
  \bibfield{journal}{%
  \Doi{10.1016/S0010-4655(03)00222-4}{\bibinfo {journal}
  {Comput.Phys.Commun.}}\ }%
  \textbf{\bibinfo {volume} {153}},\ \bibinfo {pages} {275} (\bibinfo {year}
  {2003}),\ \Eprint{http://arxiv.org/abs/hep-ph/0301101}{arXiv:hep-ph/0301101}%
  \bibAnnoteFile{NoStop}{Porod:2003um}%
\bibitem{Porod:2011nf}%
  \BibitemOpen
  \bibfield{author}{%
  \bibinfo {author} {\bibfnamefont{W.}~\bibnamefont{Porod}}\ and\ \bibinfo
  {author} {\bibfnamefont{F.}~\bibnamefont{Staub}},\ }%
  \bibfield{journal}{%
  \Doi{10.1016/j.cpc.2012.05.021}{\bibinfo {journal} {Comput.Phys.Commun.}}\ }%
  \textbf{\bibinfo {volume} {183}},\ \bibinfo {pages} {2458} (\bibinfo {year}
  {2012}),\ \Eprint{http://arxiv.org/abs/1104.1573}{arXiv:1104.1573 [hep-ph]}%
  \bibAnnoteFile{NoStop}{Porod:2011nf}%
\bibitem{Degrande:2011ua}%
  \BibitemOpen
  \bibfield{author}{%
  \bibinfo {author} {\bibfnamefont{C.}~\bibnamefont{Degrande}}, \bibinfo
  {author} {\bibfnamefont{C.}~\bibnamefont{Duhr}}, \bibinfo {author}
  {\bibfnamefont{B.}~\bibnamefont{Fuks}}, \bibinfo {author}
  {\bibfnamefont{D.}~\bibnamefont{Grellscheid}}, \bibinfo {author}
  {\bibfnamefont{O.}~\bibnamefont{Mattelaer}}, \emph{et~al.},\ }%
  \bibfield{journal}{%
  \Doi{10.1016/j.cpc.2012.01.022}{\bibinfo {journal} {Comput.Phys.Commun.}}\ }%
  \textbf{\bibinfo {volume} {183}},\ \bibinfo {pages} {1201} (\bibinfo {year}
  {2012}),\ \Eprint{http://arxiv.org/abs/1108.2040}{arXiv:1108.2040 [hep-ph]}%
  \bibAnnoteFile{NoStop}{Degrande:2011ua}%
\bibitem{Alwall:2014hca}%
  \BibitemOpen
  \bibfield{author}{%
  \bibinfo {author} {\bibfnamefont{J.}~\bibnamefont{Alwall}}, \bibinfo {author}
  {\bibfnamefont{R.}~\bibnamefont{Frederix}}, \bibinfo {author}
  {\bibfnamefont{S.}~\bibnamefont{Frixione}}, \bibinfo {author}
  {\bibfnamefont{V.}~\bibnamefont{Hirschi}}, \bibinfo {author}
  {\bibfnamefont{F.}~\bibnamefont{Maltoni}}, \emph{et~al.},\ }%
  \bibfield{journal}{%
  \Doi{10.1007/JHEP07(2014)079}{\bibinfo {journal} {JHEP}}\ }%
  \textbf{\bibinfo {volume} {1407}},\ \bibinfo {pages} {079} (\bibinfo {year}
  {2014}),\ \Eprint{http://arxiv.org/abs/1405.0301}{arXiv:1405.0301 [hep-ph]}%
  \bibAnnoteFile{NoStop}{Alwall:2014hca}%
\bibitem{CMS:2012uwa}%
  \BibitemOpen
  \bibfield{author}{%
  \bibinfo {author} {\bibnamefont{CMS-Collaboration}}}\ %
   (\bibinfo {year} {2012}),\
  \Eprint{http://arxiv.org/abs/CMS-PAS-EXO-12-017}{CMS-PAS-EXO-12-017}%
  \bibAnnoteFile{NoStop}{CMS:2012uwa}%
\end{thebibliography}
\end{document}